\documentclass[prd,superscriptaddress,twocolumn,showpacs]{revtex4}
\pdfoutput=1 

\usepackage{amsmath}
\usepackage{amsfonts}
\usepackage{amssymb}
\usepackage{amsthm}
\usepackage{epsfig}
\usepackage{graphicx}
\usepackage[dvips]{hyperref}
\usepackage{color}

\newcommand{\dmsq}{\Delta m^2}
\newcommand{\delCP}{\delta_{\rm CP}}
\newcommand{\sinsqth}{\sin^2\theta}
\newcommand{\chimin}{\chi^2_{\rm min}}
\newcommand{\beq}{\begin{equation}}
\newcommand{\eeq}{\end{equation}}
\newcommand{\beqa}{\begin{eqnarray}}
\newcommand{\eeqa}{\end{eqnarray}}
\newcommand{\barr}{\begin{array}}
\newcommand{\earr}{\end{array}}
\newcommand{\bit}{\begin{itemize}}
\newcommand{\eit}{\end{itemize}}
\newcommand{\nn}{\nonumber}
\def\gs{\mathrel{
   \rlap{\raise 0.511ex \hbox{$>$}}{\lower 0.511ex \hbox{$\sim$}}}}
\def\ls{\mathrel{
   \rlap{\raise 0.511ex \hbox{$<$}}{\lower 0.511ex \hbox{$\sim$}}}}

\begin{document}

\title{2540 km: Bimagic baseline for 
neutrino oscillation parameters} 

\author{Amol Dighe}
\affiliation{Tata Institute of Fundamental Research, 
Homi Bhabha Road, Colaba, Mumbai 400005, India}

\author{Srubabati Goswami}
\affiliation{Physical Research Laboratory, Navrangpura,
Ahmedabad 380009, India}

\author{Shamayita Ray}
\affiliation{Laboratory for Elementary-Particle Physics, 
Cornell University, Ithaca, NY 14853, USA}

\date{\today}

\begin{abstract}
We show that a source-to-detector distance of 2540 km, 
motivated recently \cite{2540-umasankar} for a narrow band superbeam,
offers multiple advantages for a low energy neutrino factory 
with a detector that can identify muon charge.
At this baseline, for any neutrino hierarchy, the wrong-sign 
muon signal is almost independent of CP violation and $\theta_{13}$ 
in certain energy ranges. 
This allows the identification of the hierarchy in a clean way.
In addition, part of the muon spectrum is also sensitive to
the CP violating phase and $\theta_{13}$, so that the same setup can 
be used to probe these parameters as well.
\end{abstract}


\maketitle


{\it Introduction.}--- 
The data from ongoing neutrino experiments confirm that neutrinos
have distinct masses $m_1, m_2, m_3$ 
and the three neutrino flavors 
$\nu_e, \nu_\mu, \nu_\tau$ mix among themselves.
While the mass squared difference $\Delta m^2_{21} \equiv m_2^2 - m_1^2$
and the magnitude of $\Delta m^2_{31} \equiv m_3^2 -m_1^2$,
as well as two of the mixing angles, $\theta_{12}$ and $\theta_{23}$,
are well measured, three 
parameters of the leptonic mixing matrix 
still remain elusive: the mixing angle $\theta_{13}$, the sign of 
$\Delta m^2_{31}$, and the CP phase $\delCP$ \cite{latest-fit}.
The determination of hierarchy (NH/normal: $\Delta m^2_{31} >0$,
IH/inverted: $\Delta m^2_{31} <0$), in particular, would be crucial in
identifying the mechanism of neutrino mass generation \cite{muchun}.

If the actual value of $\theta_{13}$ is not much below the
current $3\sigma$ bound of $\theta_{13} < 12^\circ$, it may be
measured at detectors at a distance of $\ls 1$ km from a reactor/accelerator.
In order to determine the hierarchy, however, the most efficient avenue
is to have the neutrinos travel through Earth for thousands of km
before detection. Here, the difference between Earth matter effects  
in the two hierarchies can help in distinguishing them.
This can be achieved,  for instance, by using the decay of accelerated muons --
$\mu^+$ or $\mu^-$ -- as a source (``neutrino factory'' (NF)) and
a detector that can detect muons and identify them as 
right-sign (the same sign as the source) or wrong-sign.

The wrong-sign muon signal is hailed as the ``golden channel'' 
since it is sensitive to  all the three parameters:
$\theta_{13}$, the sign of $\Delta m^2_{31}$, and $\delCP$. 
However,  the dependence on  $\delCP$  also introduces large
uncertainties, making the  unambiguous determination 
of the true parameters difficult \cite{degpapers,magic}.
A potential way out is to have the detector at 
$\sim 7500$ km (``the magic baseline'' \cite{magic,magic2}) from the source, 
where the effect of CP violation vanishes for both hierarchies. 
However, this very feature makes it impossible to measure the 
CP phase at this baseline. 
Moreover, such a long baseline requires an extremely well-collimated
muon source, else the flux at the detector is highly reduced.

It is therefore desirable to look for a shorter baseline that will
still give a wrong-sign muon signal independent of the CP phase
for one of the hierarchies, albeit only in a part of the spectrum.
The remaining part of the spectrum would still be sensitive to
the CP phase and can be used to detect CP violation for the same hierarchy. 

In the context of a $\nu_\mu$ superbeam, it was recently pointed out 
\cite{2540-umasankar} that  the baseline of 2540 km satisfies 
the above condition for IH at a neutrino energy of $3.3$ GeV and 
a narrow band neutrino beam was therefore deemed desirable.
In this Letter, we point out that this baseline 
{\it also}  satisfies the desired condition for NH, at the energy 1.9 GeV. 
The two energies at which the desired condition is satisfied are termed 
as magic energies, and the baseline is referred to as ``bimagic''. 
The bimagic property, first realized in this work, makes it  
more desirable to have a broadband beam covering the range 1--4 GeV. 
We use the $e$--$\mu$ channel in a low energy neutrino factory (LENF) 
with a muon energy of 5 GeV \cite{lownufac},
as opposed to the $\mu$--$e$ channel used for superbeams \cite{2540-umasankar}. 
The detection of muons is easier compared to that of electrons. 
Moreover with muon charge identification, 
NFs do not have beam contamination problems,
thus enabling sensitivity to smaller $\theta_{13}$ values. 
Thus, the bimagic nature in conjunction with a LENF 
helps in an efficient identification of hierarchy,
nonzero $\theta_{13}$ and CP violation, even with a single polarity
of decaying muons, as we shall motivate and demonstrate in this Letter.
It is remarkable that the distance 2540 km also happens to be close to the 
distance between Brookhaven and Homestake \cite{bnlhomestake},
as well as that between CERN and Pyhasalmi mine \cite{pyhasalmi},
which is one of the proposed sites for the LENA detector. 

\smallskip

{\it The bimagic baseline.}---
The source beam from a neutrino factory that accelerates $\mu^+$ 
consists of $\bar{\nu}_\mu$ and $\nu_e$. Charged current interactions
at the detector can give muons in two ways: the original 
$\bar\nu_\mu$ that survive as $\bar\nu_\mu$ give $\mu^+$ (right-sign
muons) while the original $\nu_e$ that oscillate to $\nu_\mu$
give $\mu^-$ (wrong-sign muons).
The oscillation probability $P_{\nu_e \to \nu_\mu}$, 
relevant for the wrong-sign muon signal,
can be written in the constant matter density approximation as  
\cite{akhmedov-prob}
\beqa
P_{e \mu}&=&4 s_{13}^2 s_{23}^2 \frac{\sin^2{[(1 -\hat A)\Delta]}}{(1-\hat A)^2} 
+ \alpha^2 \sin^2{2\theta_{12}} c_{23}^2 \frac{\sin^2{\hat A \Delta}}{\hat A^2} \nn \\ 
&& +  2 \alpha s_{13}  \sin{2\theta_{12}} \sin{2\theta_{23}} \cos{(\Delta - \delCP)} 
\times \nn \\
&& \phantom{leave some space}
\frac{\sin{\hat A \Delta}}{\hat A} \frac{\sin{[(1-\hat A)\Delta]}}{(1-\hat A)} \; ,
\label{P-emu}  
\eeqa
keeping terms up to second order in 
$\alpha \equiv \dmsq_{21}/\dmsq_{31}$ and $s_{13}$.
Here $s_{ij} \equiv \sin{\theta_{ij}}$, $c_{ij} \equiv \cos{\theta_{ij}}$.  Also,
\beqa
\hat A \equiv 2\sqrt{2} G_F n_e E_\nu / \dmsq_{31} \; , \quad 
\Delta \equiv \dmsq_{31} L / (4 E_\nu) \; ,
\eeqa
where $G_F$ is the Fermi constant and $n_e$ is the electron number density.
For neutrinos, the signs of $\hat{A}$ and $\Delta$ are positive for normal 
hierarchy and negative for inverted hierarchy. $\hat{A}$ picks up an extra 
negative sign for anti-neutrinos. 
The last term in Eq.~(\ref{P-emu}) clearly mixes the dependence on
hierarchy and $\delCP$, leading to a degeneracy between them 
\cite{degpapers}, 
which can be overcome if one manages to have either 
$\sin(\hat A \Delta) = 0$ or $\sin[(1-\hat A)\Delta] = 0$. 
The first condition is achieved at the magic baseline ($L \sim 7500$ km) 
for all $E_\nu$ and for both the hierarchies. 
The second condition, on the other hand, is sensitive to hierarchy. 
This sensitivity can be maximized if one has $\sin[(1-\hat A)\Delta] = 0$ 
for one of the hierarchies and $\sin[(1-\hat A)\Delta] = \pm 1$ for the other.   
In such a situation,
only the ${\cal{O}}(\alpha^2)$ term in Eq.~(\ref{P-emu}) survives
for the hierarchy for which $\sin[(1-\hat A)\Delta] = 0$, 
making $P_{e\mu}$ independent of both $\delCP$ and $\theta_{13}$. 
At the same time, for the other hierarchy the
first term in Eq.~(\ref{P-emu}) enhances the number of
events as well as $\theta_{13}$ sensitivity, and the third 
term enhances the sensitivity to $\delCP$. 

If we demand ``IH-noCP'' (no sensitivity to CP phase in IH), these
conditions imply
\begin{subequations}
\beqa
(1+ |\hat A|) \cdot |\Delta| & = &  n \pi \; \quad \quad \quad \quad \quad
{\text{for IH}} \; ,  
\label{ih-0} \\
(1- |\hat A|)\cdot |\Delta| & = &  (m -1/2) \pi \; \quad {\text{for NH}} \; ,
\label{nh-1}
\eeqa
\label{ih-magic}
\end{subequations}
where $n,m$ are integers, $n > 0$.
These two conditions are exactly satisfied at a particular baseline 
and energy, given by
\begin{subequations}
\beqa
\rho L {\rm (km \ g/cc)} & \approx & (n-m+1/2) \times  16300 \; , 
\label{rhol-ihnocp} \\
E_\nu {\rm (GeV)} & = & \frac{4}{5} 
\frac{\Delta m^2_{31} ({\rm eV}^2) \, L {\rm (km)}} {(n+m-1/2)} \; .
\label{enu-ihnocp}
\eeqa
\end{subequations}
Note that the relevant $L$ is independent of any oscillation 
parameters. A viable solution for these set of equations (with $n=1$ and $m=1$) is
$L \approx 2540$ km, $\rho=3.2$ g/cc and $E_\nu \equiv E_{\rm IH} \approx 3.3$ GeV,
as was first pointed out in \cite{2540-umasankar}.
On the other hand, one may demand ``NH-noCP'' (no sensitivity to CP
phase in NH), which leads to the conditions
\begin{subequations}
\beqa
(1 - |\hat A|)\cdot |\Delta| & = &  n \pi \; \quad \quad \quad \quad \quad
{\text{for NH}} \; ,  
\label{nh-0} \\
(1 + |\hat A|) \cdot |\Delta| & = &  (m -1/2) \pi \; \quad {\text{for IH}} \; ,
\label{ih-1}
\eeqa
\label{nh-magic}
\end{subequations}
with $n,m$ integers, $n \neq 0$ and $m>0$.
These lead to the same condition on $L$ as in Eq.~(\ref{rhol-ihnocp})
except for an overall negative sign, while $E_\nu$ continues to be given
by Eq.~(\ref{enu-ihnocp}).
These conditions are also satisfied at $L=2540$ km
(for $n=1$ and $m=2$) at $E_\nu \equiv E_{\rm NH} \approx 1.9$ GeV.
The magic energies $E_{\rm IH}$ and $E_{\rm NH}$ would be suitable for a
neutrino factory with a parent muon energy of $\sim 5$ GeV.

Eqs.~ (\ref{rhol-ihnocp}, \ref{enu-ihnocp})  
indicate that many combinations of $n$ and $m$ are possible
for a given baseline. Indeed, the 2540 km baseline also satisfies 
IH-noCP at $E_{\rm IH2} \approx 1.3$ GeV ($n=2, m=2$) and 
NH-noCP at $E_{\rm NH2} \approx 0.9$ GeV ($n=2, m=3$). 
However the flux at these energies would be small,
so we do not consider these in this Letter.

\begin{figure}[t]
\begin{center}
\includegraphics[width=8.2cm,height=4.5cm]{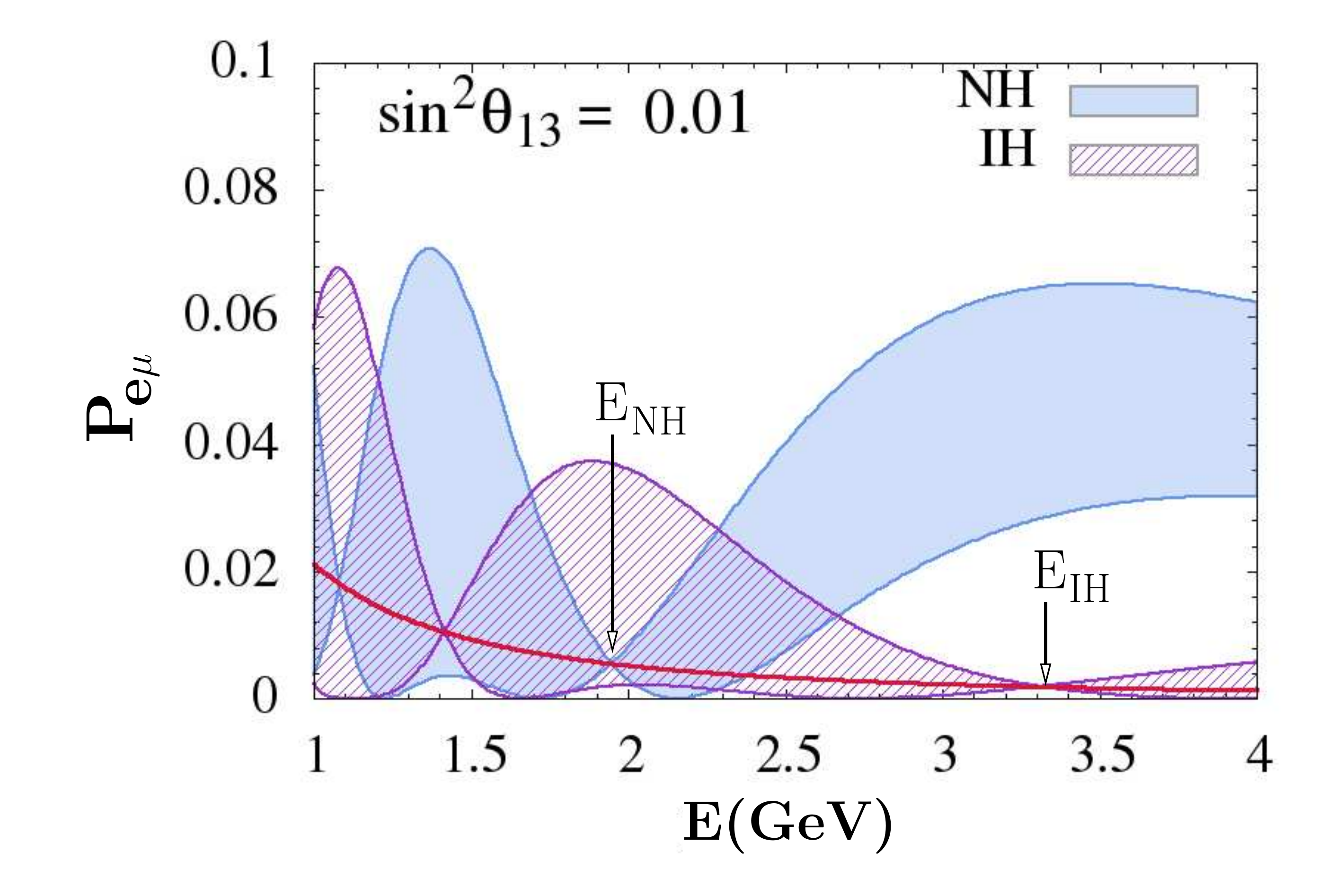}
\caption{Conversion probability $P_{e \mu}$ for $L = 2540$ km. 
The bands correspond to $\delCP \in (0, 2\pi)$. Other 
parameters are taken as $\dmsq_{21}=7.65\times 10^{-5}$ eV$^2$, 
$|\dmsq_{31}|=0.0024$ eV$^2$, $\sin^2\theta_{12}=0.3$ and 
$\sin^2\theta_{12}=0.5$. 
The red (solid) line corresponds to $\theta_{13} = 0$.
\label{fig:prob}}
\end{center}
\end{figure}

Fig.~\ref{fig:prob} shows the probability $P_{e \mu}$ 
for $\sin^2\theta_{13}=0, 0.01$.  
In this and all other plots,
we have solved the exact neutrino propagation equation numerically 
using the Preliminary Reference Earth Model \cite{prem}. 
Clearly the IH-noCP and NH-noCP conditions are satisfied at
the energies $E_{\rm IH}$ and $E_{\rm NH}$, respectively. 
At $E_{\rm IH}$, the probabilities $P_{e\mu}$ for NH and IH are
distinct, hence a measurement of the neutrino spectrum
around this energy would be a clean way of distinguishing between
the hierarchies.
The oscillatory nature of $P_{e\mu}$ for non-zero $\theta_{13}$ vis-a-vis the
monotonic behavior for $\theta_{13}=0$  helps in the discovery of
a nonzero $\theta_{13}$.
Finally, the significant widths of the bands (near $E_{\rm IH}$ for NH, 
and near $E_{\rm NH}$ for IH) imply sensitivity to $\delCP$.

The simplified forms of probabilities at the magic energies
offer insights into the CP sensitivity at this baseline.
At $E_{\rm IH}$, we have
\beqa
P_{e\mu}({\rm IH}) & \approx & 18 \alpha^2 s_{12}^2 c_{12}^2 c_{23}^2 \; , \nn \\
P_{e \mu}({\rm NH}) & \approx & 18 \alpha^2 s_{12}^2 c_{12}^2 c_{23}^2 
+ 9 s_{13}^2 s_{23}^2 \nn \\
& & \hspace{-1.5cm} -  18 \sqrt{2} \alpha s_{12} c_{12} s_{23} c_{23} s_{13} 
\cos(\delCP + \pi/4) \; ,
\label{Eih-prob}
\eeqa
while at $E_{\rm NH}$, we have
\beqa
P_{e\mu}({\rm NH}) & \approx & 50 \alpha^2 s_{12}^2 c_{12}^2 c_{23}^2 \; , \nn \\
P_{e \mu}({\rm IH}) & \approx & 50 \alpha^2 s_{12}^2 c_{12}^2 c_{23}^2 
+ (25/9)  s_{13}^2 s_{23}^2 \nn \\
& & \hspace{-1.5cm} -  (50 \sqrt{2}/3) \alpha s_{12} c_{12} s_{23} c_{23} s_{13} 
\cos(\delCP + \pi/4) \; .
\label{Enh-prob}
\eeqa
Near the magic energies, where the CP sensitivity is the
highest, the $\delCP$ values giving the highest and the lowest
probabilities would be $3\pi/4$ and $7\pi/4$, respectively.


\smallskip

\begin{figure}
\includegraphics[width=8.0cm,height=3.6cm]{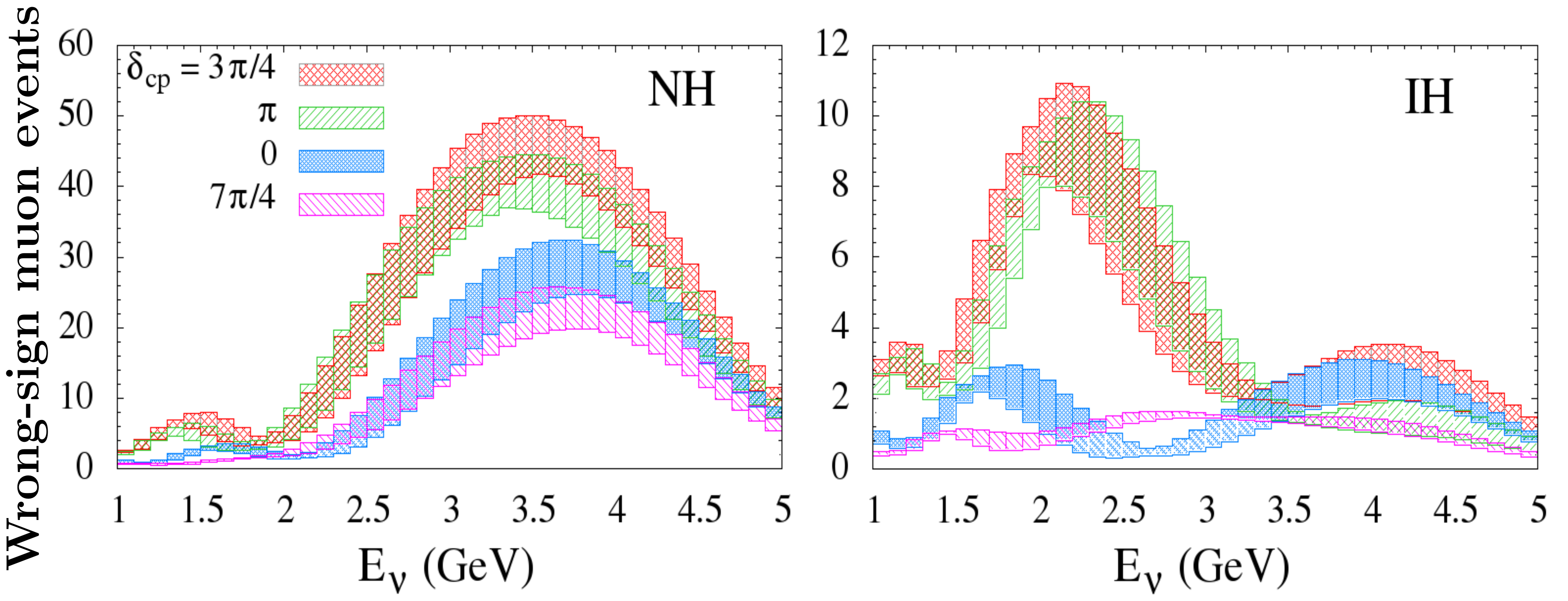}
\includegraphics[width=8.2cm,height=3.6cm]{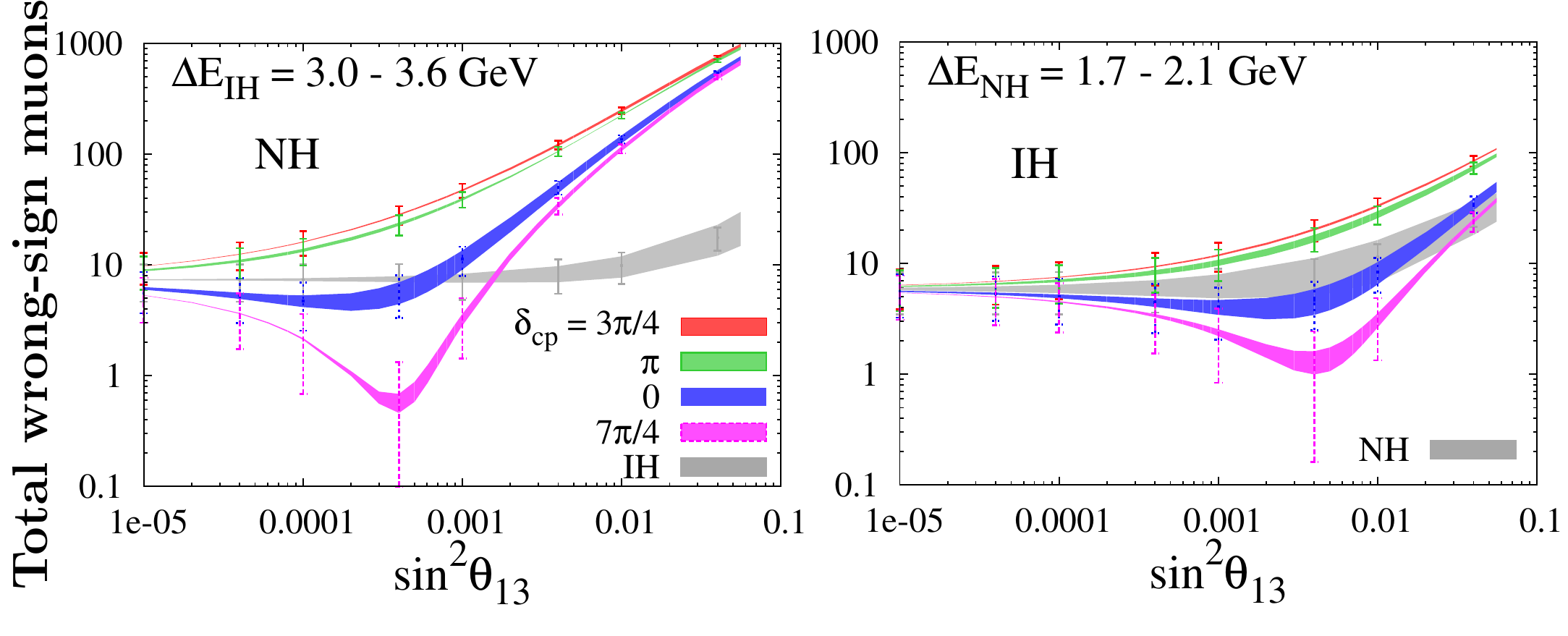}
\caption{Top panel: energy spectra of wrong-sign muons for NH (left) and
IH(right). Bottom panel: number of events for 1 year 
run, 
in the bins $\Delta E_{\rm IH}$
and $\Delta E_{\rm NH}$ as a function of $\theta_{13}$.
The bands correspond to 5\% error in $\dmsq_{31}$.
The bands in the top panels also include a 10\% error around  
$\sin^2\theta_{13} = 0.01$. 
\label{fig:events}}
\end{figure}

\smallskip

{\it Experimental setup and numerical simulation.}---
We use a magnetized totally active scintillator detector (TASD) 
which is 
generally used in the context of a  LENF  
\cite{lownufac}. 
We use a 25 kt detector with a energy threshold of 1 GeV. 
We choose a typical Neutrino factory setup with 5 GeV parent muon energy and 
$5 \times 10^{21}$ useful muon decays per year
\cite{huber, lenf-improve}. 
We consider the running with only one polarity $\mu^+$ of the
parent muon, so that 
we have a neutrino flux consisting of $\bar \nu_\mu$ and $\nu_e$. 
We assume a muon detection efficiency of 94\% for energies above 1 GeV, 
10\% energy resolution for the whole energy range up to 5 GeV and 
a background level of $10^{-3}$ for the $\nu_e \to \nu_\mu$ and 
$\bar \nu_\mu \to \bar \nu_\mu$ channels. 
Detection of $\nu_e$ or $\bar \nu_e$ is not considered in this study, 
which seems to have 
a very small effect when the initial flux is as large as above \cite{lenf-improve}. 
A 2.5\% normalization error and 0.01\% calibration error, both for 
signal and background, have also been taken into account 
throughout this study. 
The detector characteristics have been simulated by GLoBES \cite{globes}.

The top panel in Fig.~\ref{fig:events} shows the energy spectra 
of wrong-sign muon events. 
For illustration, in addition to $\delCP=0, \pi$,
we choose $\delCP = 3 \pi/4, 7\pi/4$ 
which would give the maximum $\delCP$ dependence near the
magic energies, as indicated by Eqs.~(\ref{Eih-prob}) and (\ref{Enh-prob}).
It is clear  from this figure that there is considerable sensitivity to 
CP phase  near $E_{\rm IH} (E_{\rm NH})  \approx 3.3 (1.9)$ GeV for NH (IH).  
It may be noted from the figure that the CP sensitivity for IH 
is actually better at slightly higher energies than $E_{\rm NH}$. 
This is because the $\nu_e$ spectrum at the
source as well as the cross section of $\nu_\mu$ at the detector
are strongly increasing functions of energy around $E_\nu \sim 2$ GeV, 
and push the peak in the IH spectrum to higher energies.

In order to illustrate the effectiveness of the magic energies,
we show in the bottom panel of Fig.~\ref{fig:events} 
the total number of events in two bins near the magic energies 
-- $\Delta E_{\rm IH}$ (3.0--3.6 GeV) and $\Delta E_{\rm NH}$ (1.7--2.1 GeV) --
as functions of $\theta_{13}$. 
Clearly, the bin $\Delta E_{\rm IH}$ itself is enough to identify the 
hierarchy as long as $\sin^2 \theta_{13} \gtrsim 10^{-3}$.
If the actual hierarchy is NH, this bin is also sensitive to
$\delCP$. 
The sensitivity to $\theta_{13}$ may be estimated by comparing the
error bars at different $\theta_{13}$ values.
If the actual hierarchy is IH, one needs the events data from
the energy bin $\Delta E_{\rm NH}$ in order to discern $\delCP$ and
$\theta_{13}$.
The actual identification of hierarchy and the measurement of 
$\delCP$ and $\theta_{13}$ is done by using the complete
wrong-sign events spectrum as well as the right-sign events spectrum.
We present below the results of this analysis.


\smallskip

\begin{figure}[t!]
\begin{center}
\includegraphics[width=8.0cm, height=4.0cm]{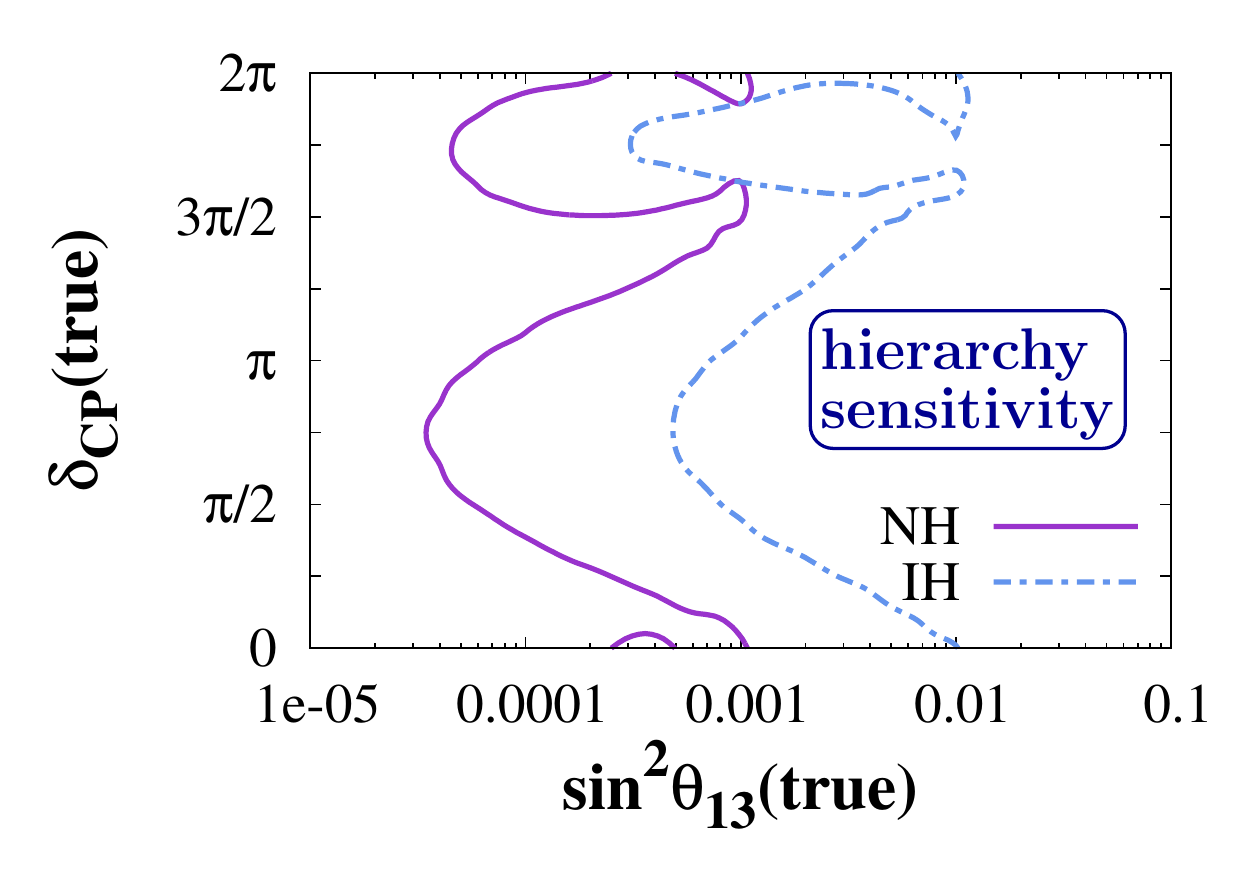}
\caption{The 3$\sigma$ hierarchy sensitivity contours,
obtained with a flux of $5 \times 10^{21}$ positive muons/year on 
a 25 kt TASD for 2.5 years, for true hierarchies as indicated.
For parameters to the right  of the contours, hierarchy can be
identified.
\label{hierarchy}}
\end{center}
\end{figure}

{\it Mass hierarchy determination.}---
In Fig.~\ref{hierarchy} we quantify the hierarchy sensitivity of the 
bimagic neutrino factory setup. 
The experimental data are generated with the chosen true hierarchy. 
The true values of $\sinsqth_{13}$ and $\delCP$ are plotted along 
the axes while the true values of the other parameters are set to the 
values quoted in Fig.~\ref{fig:prob}. 
For each pair of $\sinsqth_{13}({\rm true})$--$\delCP({\rm true})$, 
we obtain $\chimin$  by marginalizing over other parameters. 
We have taken 4\% error on each of $\dmsq_{21}$ and $\theta_{12}$, 
and 5\% error on each of $\theta_{23}$  and $\dmsq_{31}$,
for calculating the priors. 
$\delCP$ has been varied over $(0, 2\pi)$. 
A 2\% error has also been 
considered on the earth matter profile and marginalized over.

The contours in Fig.~\ref{hierarchy} suggest that if the true hierarchy 
is NH, then for favorable values of $\delCP$, 
an exposure of $\approx 3 \times 10^{23}$ muons$\times$kt may determine
the hierarchy at  3$\sigma$ even for 
$\sinsqth_{13}$ as small as $\sim 3 \times  10^{-5}$.
If the true hierarchy is IH then that can be established at 3$\sigma$  
for $\sin^2\theta_{13} \gtrsim 3 \times 10^{-4}$.
This sensitivity is better than that indicated by the superbeam studies
at this baseline \cite{bnlhomestake,2540-umasankar}.

An optimized LENF setup with a baseline of 1300 km 
has been recently proposed \cite{lenf-improve}.
However, the relatively small baseline does not allow matter effects 
to develop sufficiently, and one does not have the advantage of
the magic energies.
So the sensitivity of this setup to the hierarchy is rather limited. 
Indeed if the true hierarchy is NH, the bimagic baseline will 
rule out IH at 3$\sigma$ for $\sin^2\theta_{13}$ values almost 
an order of magnitude smaller than the expected reach of the 1300 km 
setup with the same exposure.
If IH is the true hierarchy, the performance of both the setups is 
almost the same. 
Thus the bimagic baseline is a more optimal setup as far as
the hierarchy is concerned.


\begin{figure}[t!]
\begin{center}
\includegraphics[width=8.0cm, height=7.2cm]{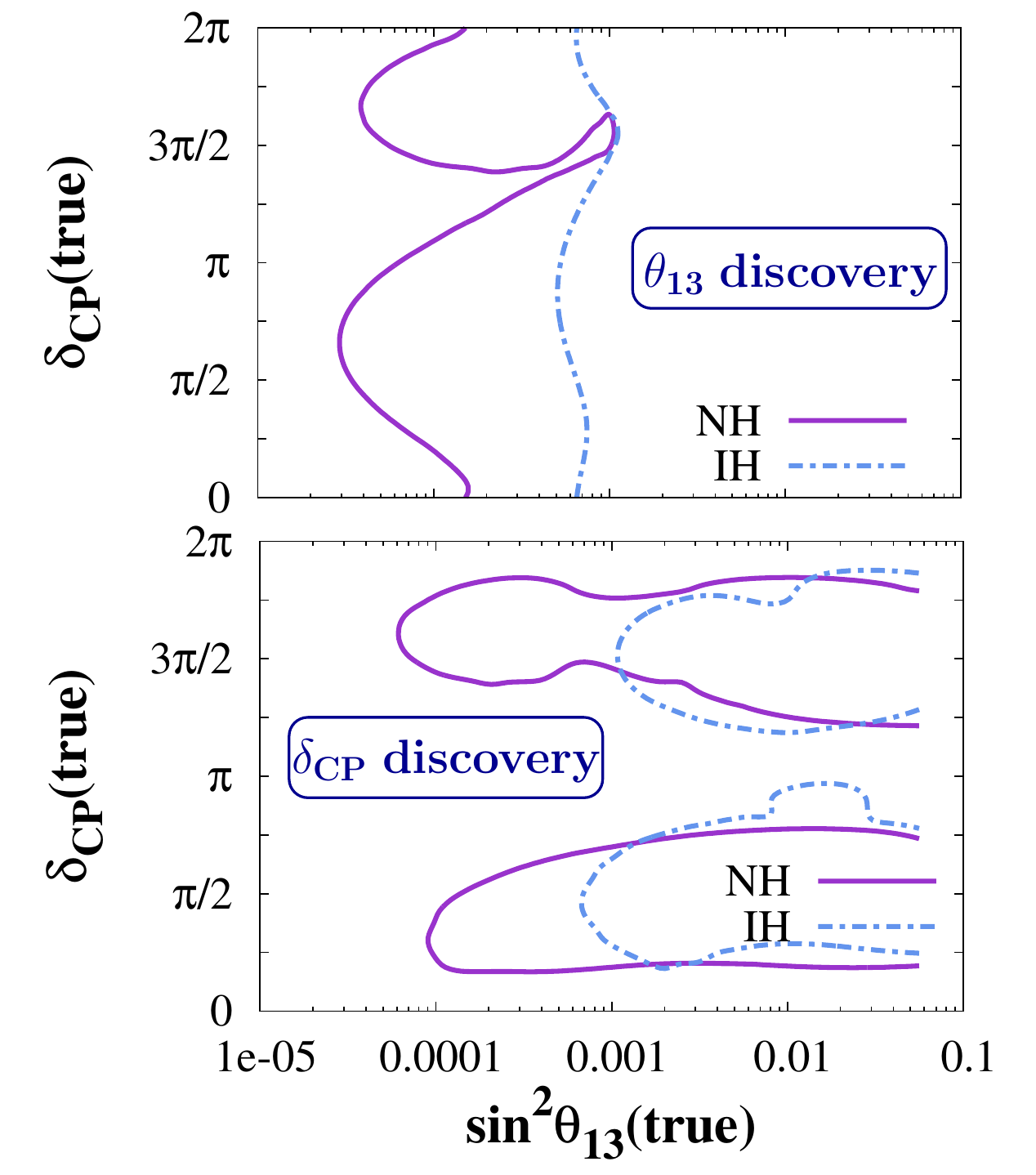}
\caption{The 3$\sigma$ discovery contours for $\theta_{13}$ 
(upper panel) and CP violating phase $\delCP$ (lower panel),
obtained with a flux of $5 \times 10^{21}$ positive muons/year on 
a 25 kt TASD for 2.5 years, for true hierarchies as indicated.
For parameters to the right  of the contours, 
the discovery of the relevant parameter is possible.
\label{th13-sensitivity}}
\end{center}
\end{figure}

\smallskip

{\it $\theta_{13}$ and $\delCP$ measurement.}---
The top panel of Fig.~\ref{th13-sensitivity} shows that 
the exposure of $\approx 3 \times 10^{23}$ muons$\times$kt
will be able to discover a nonzero $\theta_{13}$ to 3$\sigma$
as long as $\sin^2 \theta_{13} \gtrsim 10^{-3}$ for either
hierarchy and for any $\delCP$ value. For NH and
$\delCP \approx 3\pi/4$, the discovery of $\theta_{13}$ is
possible even for $\sin^2\theta_{13}$ as low as $3\times 10^{-5}$.

The bottom panel of Fig.~\ref{th13-sensitivity} shows the $\delCP$ 
discovery reach with this setup. 
It shows that the exposure allows the discovery of nonzero 
$\delCP$ for NH for $\sin^2\theta_{13}$ as low as $10^{-4}$,
as long as $\delCP \approx 3\pi/4$. 
This is the $\delCP$ value at which we expect the highest
deviation in the events spectrum from $\delCP=0$, as indicated by 
Eqs.~(\ref{Eih-prob}) and (\ref{Enh-prob}).
For IH, the results are about one order of magnitude worse than
those for NH.

The discovery potential for $\theta_{13}$ and $\delCP$ at
the bimagic baseline is comparable to that of 
the 1300 km setup if the true hierarchy is NH, 
while it is not as good if the true hierarchy is IH.
However, note that this is valid if only $\mu^+$ are available
at the source. With both polarities available, the bimagic baseline
would be almost as good as the 1300 km setup for $\theta_{13}$
and $\delCP$, and will have a better sensitivity to the hierarchy.
Indeed, once the hierarchy is identified -- for which the bimagic
baseline performs better -- running the bimagic setup
with $\mu^+$ ($\mu^-$) as the source beam for NH (IH) would offer 
a sensitivity similar to the 1300 km setup.
Thus, overall the bimagic baseline seems like an optimal one to
probe the three most important unknown parameters of the leptonic 
mixing matrix: $\theta_{13}, \delCP$ and the sign of $\Delta m^2_{31}$.

\smallskip

{\it Conclusion.}---
We have shown the ``bimagic'' nature of the 2540 km baseline: at this 
baseline with judicious choice of energies, the dependence of 
the wrong-sign muon signal on $\theta_{13}$ and $\delCP$ can be 
made to vanish for either hierarchy. This energy turns out to be around 
3.3 GeV for IH and 1.9 GeV for NH. 
This helps in an efficient identification of hierarchy even at very 
low $\theta_{13}$, when one uses a neutrino factory with
parent muon energy $\sim 5$ GeV as a source.
On the other hand the sensitivity to $\theta_{13}$ and $\delCP$ 
is maximum at $\sim$ 3.3 GeV for NH and 1.9 GeV for IH, allowing the
determination of these parameters as well with the same beam-baseline setup.
To exploit these features, a broadband beam of  a neutrino factory 
is  more effective as compared to a narrow band beam.

\smallskip

{\it Acknowledgments.}---
We thank  P. Ghoshal, J. Kopp, S. Uma Sankar  and W. Winter for useful discussions.
S.R. would like to thank Prof. Yuval Grossman 
for support and hospitality.


\begin{thebibliography}{99}

\bibitem{2540-umasankar}
S.~K.~Raut, R.~S.~Singh and S.~U.~Sankar,
  arXiv:0908.3741 [hep-ph].

\bibitem{latest-fit} 
  T.~Schwetz, M.~A.~Tortola and J.~W.~F.~Valle,
  New J.\ Phys.\  {\bf 10}, 113011 (2008)
  [arXiv:0808.2016 [hep-ph]].

\bibitem{muchun}
  C.~H.~Albright and M.~C.~Chen,
  Phys.\ Rev.\  D {\bf 74}, 113006 (2006).

\bibitem{degpapers}
  G.~L.~Fogli and E.~Lisi,
  Phys.\ Rev.\ D {\bf 54}, 3667 (1996);
%
  J.~Burguet-Castell, M.~B.~Gavela, J.~J.~Gomez-Cadenas, P.~Hernandez and O.~Mena,
  Nucl.\ Phys.\ B {\bf 608}, 301 (2001);
%
  H.~Minakata and H.~Nunokawa,
  JHEP {\bf 0110}, 001 (2001).

\bibitem{magic}
V.~Barger, D.~Marfatia and K.~Whisnant,
  Phys.\ Rev.\  D {\bf 65}, 073023 (2002).

\bibitem{magic2} 
P.~Huber and W.~Winter,
  Phys.\ Rev.\  D {\bf 68}, 037301 (2003).

\bibitem{lownufac}
S.~Geer, O.~Mena and S.~Pascoli,
  Phys.\ Rev.\  D {\bf 75}, 093001 (2007).

\bibitem{bnlhomestake}
  M.~V.~Diwan {\it et al.},
  Phys.\ Rev.\  D {\bf 68}, 012002 (2003).

\bibitem{pyhasalmi}
  J.~Peltoniemi,
  arXiv:0911.4876 [hep-ex].

\bibitem{akhmedov-prob}
E.~K.~Akhmedov, R.~Johansson, M.~Lindner, T.~Ohlsson and T.~Schwetz,
  JHEP {\bf 0404}, 078 (2004);
%
 M.~Freund,
 Phys.\ Rev.\  D {\bf 64}, 053003 (2001);
%
 A.~Cervera, A.~Donini, M.~B.~Gavela, J.~J.~Gomez Cadenas, P.~Hernandez, O.~Mena and S.~Rigolin,
 Nucl.\ Phys.\  B {\bf 579}, 17 (2000)
 [Erratum-ibid.\  B {\bf 593}, 731 (2001)].

\bibitem{prem}
A.~M.~Dziewonski and D.~L.~Anderson, 
Phys.\ Earth\ Planet.\ Inter. {\bf 25}, 297 (1981).

\bibitem{huber}
P.~Huber and T.~Schwetz,
  Phys.\ Lett.\  B {\bf 669}, 294 (2008).

\bibitem{globes}
 P.~Huber, M.~Lindner and W.~Winter,
  Comput.\ Phys.\ Commun.\  {\bf 167}, 195 (2005);
%
  P.~Huber, J.~Kopp, M.~Lindner, M.~Rolinec and W.~Winter,
  Comput.\ Phys.\ Commun.\  {\bf 177}, 432 (2007).


\bibitem{lenf-improve} 
  E.~Fernandez Martinez, T.~Li, S.~Pascoli and O.~Mena,
  Phys.\ Rev.\  D {\bf 81}, 073010 (2010).


\end{thebibliography}
\end{document}